\documentclass{article}

\usepackage[preprint]{neurips_2026}

\usepackage[utf8]{inputenc}
\usepackage[T1]{fontenc}
\usepackage{hyperref}
\usepackage{url}
\usepackage{booktabs}
\usepackage{amsfonts}
\usepackage{amsmath}
\usepackage{nicefrac}
\usepackage{microtype}
\usepackage{xcolor}
\usepackage{graphicx}
\usepackage{multirow}
\usepackage{makecell}
\usepackage{float}     
\usepackage{longtable} 

\graphicspath{{figures/}}

\title{Topic Is Not Agenda: A Citation-Community Audit of Text Embeddings}

\author{%
  Junseon Yoo \\
  Pluto Labs \\
  \texttt{junseon@pluto.im}
}

\begin{document}

\maketitle

\begin{abstract}
Vector search and retrieval-augmented generation (RAG) rest on
the assumption that cosine similarity between text embeddings
reflects conceptual relatedness. We measure where this
assumption breaks. We build an augmented citation graph over
3.58M scientific papers and partition it via Leiden CPM at two
granularities: sub-field (L1) and research-agenda (L2,
hierarchical inside each L1). Four state-of-the-art embeddings
(Gemini, Qwen3-8B, Qwen3-0.6B, SPECTER2) clear the L1 bar
reasonably (45--52\% top-10 same-rate) but stop working at L2:
only 15--21\% of top-10 neighbors share the query's research
agenda. In absolute terms, 8 of every 10 retrieved papers are
off-agenda. The failure is universal across eight scientific
domains and all four models; SPECTER2, despite its
citation-based contrastive training, is the weakest. As a
diagnostic probe, we test whether the same augmented graph
also functions as a retrieval signal: a deliberately simple
citation-count rerank reaches 57.7\% top-1 L2 on top of
LLM-expanded Boolean retrieval and 59.6\% on top of plain
BM25, on 80 curated agenda queries -- about 9 points above the
best cosine retriever (Gemini, 50.6\%) and 20 points above
BM25 alone (39.3\%). The probe isolates a slice of the
agenda-matching signal the graph carries but the embeddings
miss, connecting recent theoretical limits on single-vector
retrieval to a concrete failure mode of scientific RAG.
\end{abstract}

\section{Introduction}
\label{sec:intro}

Every deployed vector database -- Pinecone, Weaviate, Milvus,
Qdrant, Chroma, FAISS-backed stacks -- treats high cosine
similarity between text embeddings as a proxy for conceptual
relatedness. Recent theoretical work questions this proxy.
\citet{steck2024cosine} show that the cosine of learned embeddings
inherits implicit training-time scalings and is not an intrinsic
semantic quantity; \citet{weller2026theoretical} prove that for
any fixed embedding dimension there exist query--document
relevance structures no single-vector retriever can represent.
What is missing is an empirical measurement of where this proxy
breaks in a high-stakes retrieval domain and at what granularity.

Scientific literature is a good setting. Each paper comes with a
verified structural signal -- its references -- which encode expert
judgment about which prior work it builds on. We use this signal
to build two citation partitions of a 3.58M-paper augmented
citation graph: a sub-field level (L1, 73K Leiden-CPM communities)
and a finer research-agenda level (L2, 329K communities,
hierarchical inside each L1). We then ask, for each query paper,
whether its top-$k$ text-embedding neighbors share its community at
each level, for four state-of-the-art embeddings (Gemini,
Qwen3-8B, Qwen3-0.6B, SPECTER2) over eight scientific domains.

\paragraph{Contributions.}
The main contributions of this paper are:
\begin{itemize}
  \item A two-level citation partition of a 3.58M-paper augmented
  graph (153M edges) at sub-field (L1, 73K clusters) and
  research-agenda (L2, 329K clusters) granularities, validated by a
  manual read of 16 representative queries.
  \item A measurement on 1.6M papers across four state-of-the-art
  embeddings and eight scientific domains showing that top-10
  same-rate is 45--52\% at L1 but only 15--21\% at L2, with the
  pattern stable across domains, neighbor ranks, and models.
  Scaling within a family (Qwen3-0.6B $\to$ 8B) lifts L2 by only
  3--5 percentage points, and citation-based training (SPECTER2) is
  in fact the weakest of the four.
  \item A diagnostic probe on 80 curated research-agenda queries.
  Using the same augmented graph as a retrieval signal -- a
  deliberately simple citation-count rerank on top of plain BM25
  candidates -- already reaches 59.6\% top-1 L2, against 50.6\%
  for the best cosine retriever (Gemini) and 39.3\% for BM25
  alone. The probe is not a proposed retrieval method; it
  isolates the slice of the agenda-matching signal that the graph
  carries and the embeddings do not.
\end{itemize}

\paragraph{Broader implications.} Document-level reference graphs
also exist for legal opinions, patents, clinical guidelines, and
financial filings, where the same surface terms can carry different
meanings across subfields. We expect the agenda-granularity failure
to show up in these settings too: the nearest text-embedding
neighbor will often be on the same broad topic but not on the same
narrower question. As a measurement paper the direct misuse
pathway is limited; the main societal risk we see is that
publishing a specific community partition as ``ground truth'' may
reify our partitioning choices in downstream benchmarks, which we
flag in Limitations~(i).

\section{Related Work}
\label{sec:related}

\paragraph{Text embedding benchmarks and scientific embeddings.}
The Massive Text Embedding Benchmark (MTEB,
\citealt{muennighoff2023mteb}) drives progress on general-domain
retrieval, STS, and clustering. Its current leaders include Gemini
Embedding~\citep{lee2025gemini} and the Qwen3
family~\citep{zhang2025qwen3}. None of these benchmarks measure
whether a model's cosine distribution matches a domain-specific
structural graph at a controlled partition granularity. Within
scientific embeddings, SPECTER~\citep{cohan2020specter},
SPECTER2~\citep{singh2023scirepeval}, and
SciNCL~\citep{ostendorff2022scincl} use direct citations as a
contrastive training signal; one of our findings is that this
training does not transfer to cosine neighborhoods that match a
well-constructed citation partition (Section~\ref{sec:results-models}).

\paragraph{Citation-network community detection and graph-augmented retrieval.}
Bibliographic coupling~\citep{kessler1963bibliographic} and
co-citation~\citep{small1973cocitation} are the classical
multi-reference edge types we combine to build the augmented graph
(Section~\ref{sec:method-community}); Leiden~\citep{traag2019leiden}
and Louvain~\citep{blondel2008louvain} are the standard optimisers.
Most retrieval systems still rely on cosine similarity of learned
dense representations~\citep{karpukhin2020dpr}; GraphRAG
\citep{edge2024graphrag} builds a knowledge graph from text but
does not use an existing citation graph. Cross-encoder rerankers
and hybrid dense--sparse retrieval inherit the same initial
top-$k$ candidate pool, which our measurement suggests fails at
the agenda level rather than at the sub-field level. This
motivates retrieval pipelines that combine text embeddings with
the reference graph they currently ignore.

\section{Method}
\label{sec:method}

\subsection{Corpus and Sampling}
\label{sec:method-data}
Our corpus is a composite built from OpenAlex, PubMed, Semantic
Scholar, and publisher websites. The reference timestamp for all
sources is March 2026. The three indexing sources (OpenAlex,
PubMed, Semantic Scholar) each contribute title, abstract, venue,
year, and author fields for their indexed papers, together with
outgoing-reference lists. When a paper's abstract is missing across
all three indexing sources, we crawl it directly from the publisher
website. Records are deduplicated across sources by canonical DOI
and by a title\,+\,first-author\,+\,year hash.

We then restrict the resulting corpus to papers published in
SCIE-indexed journals or in CORE A*/A conferences, which yields a
pool of about 198 million papers. From this pool we select eight scientific
domains: Biology, Biomedical, Chemistry, Computer Science,
Engineering, Environmental/Earth Sciences, Materials Science, and
Physics. Seven are defined by JCR Q1 journals within the domain.
For Computer Science, where publication norms favor conferences, we
augment the Q1 set with CORE A*/A conferences. Within each domain,
we retain only original research articles (excluding reviews,
surveys, meta-analyses, editorials, and corrections), require an
abstract of at least 150 characters, and restrict to 2010 or later.
The size of each domain's filtered pool is given in
Table~\ref{tab:dataset}; the total across domains is 16.0M papers.

From each filtered domain pool we stratified-sample 500{,}000
papers by publication year, producing a 4M-paper analysis set on
which the augmented citation graph is built. We then apply three
cleaning passes to the 4M set: (i) exact-duplicate and near-duplicate
removal using canonical DOI and title--abstract hashing, which
merges preprint and published versions of the same work;
(ii) boilerplate-abstract filtering (removing publisher-template and
figure-caption-only records); and (iii) removal of papers that end
up orphans in the augmented citation graph
(Section~\ref{sec:method-community}), i.e., papers with no direct
citation and no bibliographic-coupling or co-citation tie to any
other paper in the 4M set. 10.5\% of the 4M is removed,
predominantly orphans, leaving the 3.58M-paper graph used for
community detection. From this cleaned graph we additionally
downsample to a standardized \textbf{200{,}000 papers per domain}
by publication year with a fixed random seed, giving a 1.6M-paper
analysis set. All embedding, nearest-neighbor, and neighbor-community
measurements operate on the 1.6M standardized pool; community
detection runs on the 3.58M-paper graph so that L1 and L2 community
labels are computed jointly across all eight domains.

The 500K-per-domain sample feeds the augmented graph (a denser
graph yields tighter Leiden communities); the 200K-per-domain
standardisation equalises per-domain contribution to all aggregate
statistics in Section~\ref{sec:results}. Every domain has $>200$K
papers after cleaning, so this standardisation is exact; both
sampling steps use the same fixed random seed.

\begin{table}[t]
\centering
\small
\caption{Per-domain dataset statistics.  ``Pool'' is the filtered set of JCR Q1 journals (seven domains) augmented with CORE A*/A conferences for Computer Science, restricted to original research published in 2010 or later with abstracts of at least 150 characters.  ``Sampled'' is the standardized 200{,}000-paper pool per domain used throughout our experiments.  $N_\text{L1}$ is the number of Leiden-CPM sub-field communities ($\gamma=10^{-4}$) with at least one paper in the sample; ``Max L1\%'' is the share of the pool taken by the largest single L1 community.  Baselines report the probability that two uniformly random pool papers share a community at the corresponding granularity.}
\label{tab:dataset}
\begin{tabular}{lrrrrrr}
\toprule
Domain & Pool & Sampled & $N_\text{L1}$ & Max L1\% & Base. L1 (\%) & Base. L2 (\%) \\
\midrule
Biology & 944,308 & 200,000 & 3,349 & 1.8 & 0.46 & 0.012 \\
Biomedical & 5,814,101 & 200,000 & 5,711 & 0.8 & 0.16 & 0.009 \\
Chemistry & 3,165,773 & 200,000 & 3,517 & 4.3 & 0.57 & 0.011 \\
CS & 1,114,095 & 200,000 & 3,291 & 5.1 & 1.20 & 0.028 \\
Engineering & 1,760,077 & 200,000 & 4,384 & 1.9 & 0.36 & 0.010 \\
Env./Earth & 1,169,228 & 200,000 & 3,151 & 5.2 & 1.32 & 0.024 \\
Materials & 999,817 & 200,000 & 2,897 & 2.0 & 0.67 & 0.018 \\
Physics & 1,074,586 & 200,000 & 3,432 & 11.6 & 2.63 & 0.046 \\
\midrule
\textbf{Total} & \textbf{16,041,985} & \textbf{1,600,000} & -- & -- & -- & -- \\
\bottomrule
\end{tabular}
\end{table}

\subsection{Two-Level Citation Community Detection}
\label{sec:method-community}
A text embedding matches a citation community only as well as the
community partition is itself meaningful, so the graph construction
and resolution choice below are a non-trivial piece of the method.

\paragraph{Augmented graph.}
The direct-citation subgraph over our target papers is very sparse:
each paper cites \mbox{${\sim}40$--$50$} references on average but only a
small fraction point at another target paper, leaving the induced
subgraph with mean degree $\approx 9$ and roughly 17\% of target papers
orphaned. Sparsity of this order prevents Leiden from finding
meaningful sub-field structure: at every resolution we tested, either
a single cluster swallowed the majority of the graph or the graph
atomized into two- and three-paper shards.

We instead build an \textbf{augmented citation graph} that combines
three edge sources, all computed over the full OpenAlex reference
table (${\sim}2.5$B edges, not restricted to target papers):
\begin{itemize}
  \item \emph{Direct citation}: undirected target--target edge (14.56M
        edges), unweighted with weight floor 1.0.
  \item \emph{Bibliographic coupling} (BC,
        \citealt{kessler1963bibliographic}): two target papers $a$, $b$
        that share at least three external references get an edge
        weighted by Salton cosine
        $w_\text{BC}(a, b) = |R(a)\cap R(b)| /
        \sqrt{|R(a)|\cdot|R(b)|}$, where $R(\cdot)$ is the full
        reference set. We cap ``hot'' references cited by more than 500
        target papers to remove universal-reference noise
        (132.19M edges).
  \item \emph{Co-citation} (CC, \citealt{small1973cocitation}): two
        target papers $a$, $b$ co-cited by at least three common
        citers (citers from anywhere in OpenAlex, not only target)
        get an edge weighted by Salton cosine on citer sets; citers
        with more than 200 target-paper citations are capped as
        survey-like noise (21.22M edges).
\end{itemize}
Direct edges carry weight $\geq 1$; BC and CC weights are additively
layered. When a pair of papers has edges from more than one source,
they are merged into a single edge whose weight is the sum of the
layer contributions, so the per-layer edge counts above
($14.56\text{M} + 132.19\text{M} + 21.22\text{M} = 167.97$M)
collapse to \textbf{153.18M edges over 3.58M papers} with mean
degree 85.5. The 3.58M nodes are the 89.5\% of the 4M pre-cleaning
sample that have at least one direct, BC, or CC edge to another
paper in the sample. All embedding and nearest-neighbor
measurements (Section~\ref{sec:method-neighbors}) operate on the
1.6M-paper standardized sample, which inherits L1 and L2 community
labels from this 3.58M-paper graph. BC accounts for 86\% of edges
by count; it is this multi-reference overlap signal that makes
dense community detection tractable.

\paragraph{Level 1 (sub-field): Leiden CPM.}
We run weighted undirected Leiden~\citep{traag2019leiden} on the
augmented graph, optimising the Constant Potts Model
$Q(P) = \sum_{C \in P} \bigl(e_C - \gamma \tfrac{n_C (n_C-1)}{2}\bigr)$,
where $e_C$ is the total edge weight inside community $C$ and $n_C$
its size. CPM avoids the modularity resolution limit
\citep{blondel2008louvain} and exposes one density-interpretable
parameter $\gamma$. We sweep $\gamma$ across eight values from
$10^{-6}$ to $10^{-2}$ (Appendix~\ref{app:resolution}) and choose
$\gamma_\text{L1}=10^{-4}$ for the main analysis: this gives
73{,}477 communities, with the largest covering 1.5\% of the graph
and 40 clusters above 10K papers; manual inspection confirms each
of the top-10 maps cleanly to a research sub-field. We refer to L1
communities as \emph{sub-fields}.

\paragraph{Level 2 (research agenda): hierarchical CPM.}
L1 is too coarse to separate the specific research agendas a single
lab pursues. Raising $\gamma$ globally is not a remedy: at
$\gamma = 10^{-2}$ the global partition shreds cross-sub-field
edges and produces more than 100K singletons. We instead run Leiden
CPM a second time inside each L1 community of size $\geq 200$ with
$\gamma_\text{L2}=10^{-2}$ (1{,}896 such communities are split;
smaller L1s inherit their L1 label as L2). The resulting L2
partition has 328{,}738 communities with non-singleton mean size
19 and maximum size 1{,}712; 51.2\% of papers sit in L2
communities of 10--99 papers, an agenda-scale typical of a
specific research thread.
Appendix~\ref{app:l2-validation} reports a manual read of 16
queries confirming that L2 assignments correspond to what
researchers would call a narrow research agenda (e.g., separating
UV-spectroscopy, X-ray, near-infrared Brackett-series, and optical
interferometry studies of Herbig Ae/Be stars into distinct L2s
inside a single L1 sub-field).

\subsection{Embedding Models}
\label{sec:method-embeddings}
We evaluate four text embedding models spanning commercial,
open-source, and scientific-domain-specialized families. Three of
them are general-purpose models with public
MTEB~\citep{muennighoff2023mteb} scores (overall rank as of our
snapshot in parentheses): \textbf{Qwen3-Embedding-8B}~\citep{zhang2025qwen3},
a 7.6B-parameter open-source model (rank 4); \textbf{Gemini
Embedding 001}~\citep{lee2025gemini}, a commercial API model
(rank 5); and \textbf{Qwen3-Embedding-0.6B}, the 0.6B variant of
the Qwen3 family included to separate scale from training recipe
(rank 17). The fourth, \textbf{SPECTER2}~\citep{singh2023scirepeval},
is a scientific-literature embedding model trained with
citation-based contrastive learning and widely adopted as the de
facto academic embedding; it is not on MTEB but is the standard
choice for academic retrieval.

For each model and each paper we embed the concatenation of title and
abstract joined by the model's recommended separator (the SPECTER2
tokenizer's \texttt{[SEP]} token; a newline for Gemini and the Qwen3
family) and apply the model's recommended pooling (last-token pooling
for Gemini and Qwen3, \texttt{[CLS]} pooling for SPECTER2). Embeddings
are $\ell_2$-normalized so inner-product search reduces to cosine
similarity. This yields four 1.6M-paper embedding matrices. Per-model
tokenizer, max-length, and pooling details are in
Appendix~\ref{app:embedding-config}.

\subsection{Hierarchical Neighbor Analysis}
\label{sec:method-neighbors}
For each paper $p$ in the 200K standardized pool
$\mathcal{P}_\mathcal{D}$ of domain $\mathcal{D}$ and each
embedding model $m$, we retrieve $p$'s top-100 cosine neighbors
within $\mathcal{P}_\mathcal{D}$ using FAISS~\citep{johnson2021faiss}
exact inner-product search on $\ell_2$-normalised vectors. Let
$\mathrm{nn}_k(p)$ be the $k$-th neighbor and $c_\ell(\cdot)$ the
community label at level $\ell \in \{\text{L1}, \text{L2}\}$. Our
primary observable is the rank-$k$ same-community rate
$s_k^{(\ell)}(m, \mathcal{D})
= \mathbb{E}_{p}\bigl[\mathbf{1}\{c_\ell(\mathrm{nn}_k(p))
  = c_\ell(p)\}\bigr]$
for $k \in \{2, 5, 10, 25, 50, 100\}$ (rank 1 is the query
itself). The chance baseline is
$b^{(\ell)}(\mathcal{D}) = \sum_{c} \rho_c^2$, the probability that
two uniformly random papers share an $\ell$-level community
($b^{(\text{L1})} \in [0.16\%, 2.63\%]$,
$b^{(\text{L2})} \in [0.009\%, 0.046\%]$ per domain). Throughout
the paper we highlight $k=10$; the full rank sweep is shown in
Figure~\ref{fig:rank-sweep} (Section~\ref{sec:results-rank}).

\subsection{Citation-Aware Retrieval}
\label{sec:method-retrieval}

\paragraph{Curated query construction.}
For each domain we used Claude Opus~4.7 to extract 25 candidate
research-agenda topics from highly-cited papers in that domain,
with deduplication. From these 25, we retained the 10 agendas
whose Boolean keyword search returned the largest number of
matching papers in our corpus, so that every retained agenda has
enough indexed papers to support a top-10 retrieval comparison.
The resulting 80 curated queries (10 per domain $\times$ 8 domains)
are listed in Appendix~\ref{app:queries}.

\paragraph{Retriever pipeline.}
We compare seven retrievers in two groups. \emph{Candidate-only}:
BM25 lexical retrieval (default \texttt{rank\_bm25} parameters,
top-10 over title\,+\,abstract) and four cosine embeddings (Gemini,
Qwen3-8B, Qwen3-0.6B, SPECTER2; top-10 cosine neighbors from the
1.6M pool). \emph{Citation-rerank}: a citation-graph-aware
retriever (\textbf{Graph}) and a BM25\,+\,citation rerank variant
(\textbf{BM25+cite}).

The Graph retriever runs two LLM agents (both
gemini-flash-lite~2.5, no shared state). A \emph{search-strategist}
generates 3--5 diverse Boolean queries that target different facets
of the research intent; each query is evaluated against the title
and abstract fields of the per-domain corpus, restricted to the
query's source domain and re-ranked by internal citation count
(incoming citations from other papers in the same result set),
keeping the top 10 as the sample to judge. A \emph{research-analyst}
agent scores the candidate query on a continuous relevance scale in
$[0, 1]$ based on the top-10 paper titles; a candidate passes if
its score is at least $0.8$. Among passing candidates we select the
query with the largest result set, extract its top-1{,}000 papers
by the same internal citation count, and intersect with the 1.6M
standardized pool. BM25+cite swaps the LLM-Boolean candidate
generator for plain BM25: top-1{,}000 BM25 hits are re-ranked by
the same internal citation count. We also report three Reciprocal
Rank Fusion baselines that combine BM25, Gemini, and citation
scores at the rank level (Section~\ref{sec:results-retrieval}).
The full text of both LLM prompts is in Appendix~\ref{app:prompts}.

\section{Results}
\label{sec:results}

We report five observations. Sections
\ref{sec:results-gap}--\ref{sec:results-domain} give the quantitative
$k$-NN measurements. Section~\ref{sec:results-validation} anchors them
in a manual read of concrete queries. All numbers are drawn from
Table~\ref{tab:hierarchical}. The full rank sweep and per-domain
baselines are in Appendix~\ref{app:full-tables}.

\subsection{Embeddings Fail at Agenda Granularity}
\label{sec:results-gap}
\begin{figure}[t]
  \centering
  \includegraphics[width=\linewidth]{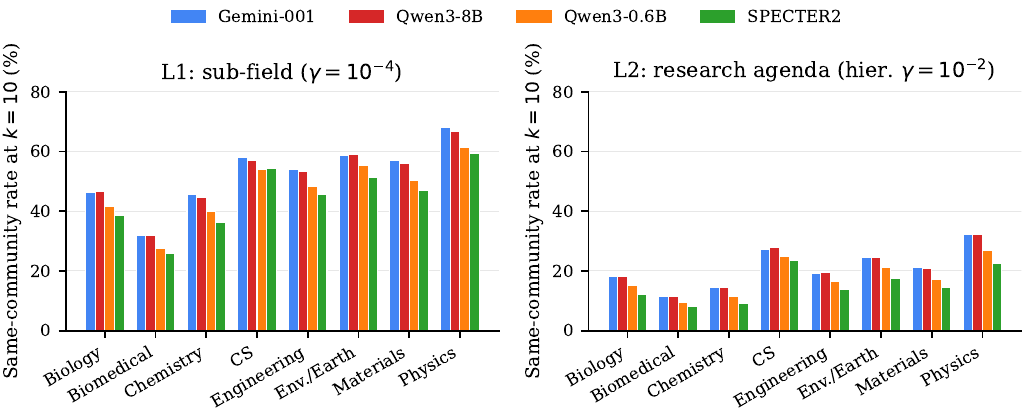}
  \caption{Top-10 same-community rate at two citation-community
  granularities across four embedding models and eight scientific
  domains. \textbf{Left (L1, sub-field):} embeddings place 44--52\% of
  the top-10 neighbors in the query's sub-field on average---roughly
  half of the neighborhood is topically on target. \textbf{Right (L2,
  research agenda):} the same neighborhoods contain the query's
  specific research agenda in only 15--21\% of slots. The
  $\sim$30-percentage-point gap between the two panels is the central
  observation of this paper. Baselines (not drawn; see
  Table~\ref{tab:dataset}) are 0.16--2.63\% at L1 and 0.009--0.046\% at
  L2, so both panels sit far above chance in absolute terms.}
  \label{fig:hierarchical}
\end{figure}

Figure~\ref{fig:hierarchical} is the central result. Both panels
plot the top-10 same-community rate for each (model, domain) cell.
The left panel uses L1 sub-field assignments (73{,}477 communities,
$\gamma_\text{L1}=10^{-4}$). The right panel uses L2 research-agenda
assignments (328{,}738 communities, hierarchical
$\gamma_\text{L2}=10^{-2}$).

The L1 panel is a sanity check on the partition: Gemini places
$52.4\%$ of its top-10 neighbors in the query's sub-field, SPECTER2
$44.7\%$, and every cell is well above chance (enrichment
23--198$\times$ over the pool baseline). Embeddings clear the
sub-field bar.

At L2, the same embeddings place only $21.0\%$ (Gemini) to
$15.2\%$ (SPECTER2) of their top-10 neighbors in the query's
community -- 8 of 10 retrieved papers belong to a different
research agenda than the query. Enrichment is large
(490--2{,}017$\times$) but only because the L2 baseline is tiny
($0.01$--$0.05\%$); what matters operationally is the absolute
15--21\%. The L1$\to$L2 drop is 19--35 percentage points per domain
and holds for every model.

\begin{table}[t]
\centering
\small
\caption{Top-10 nearest-neighbor same-community rate at two citation-community granularities. L1 (sub-field, $\gamma=10^{-4}$) partitions papers into 73{,}477 clusters (max size 53{,}498); L2 (agenda, hierarchical $\gamma=10^{-2}$) partitions them into 328{,}738 clusters (mean non-singleton size 19). Baseline is the probability that two uniformly random papers in the 200{,}000-paper pool share a community. Across (model, domain) cells, enrichment over baseline ranges from 23$\times$ to 198$\times$ at L1 (per-model means 77--92$\times$) and from 490$\times$ to 2{,}017$\times$ at L2 (per-model means 889--1{,}262$\times$). The model ordering Gemini $\approx$ Qwen3-8B $>$ Qwen3-0.6B $>$ SPECTER2 holds in every (model, domain) cell except Computer Science at L1, where SPECTER2 narrowly edges Qwen3-0.6B (54.3\% vs.\ 53.8\%). The L1 $\to$ L2 drop ranges from 19 (Biomedical) to 35 (Physics) percentage points: embeddings capture sub-field but not the narrower research agenda.}
\label{tab:hierarchical}
\begin{tabular}{lrrrrr}
\toprule
Domain & \makecell[r]{Baseline\\(\%)} & Gemini & Qwen3-8B & Qwen3-0.6B & SPECTER2 \\
\midrule
\multicolumn{6}{l}{\textit{L1 (sub-field, $\gamma=10^{-4}$)}} \\
\midrule
Biology & 0.456 & 46.4 & 46.7 & 41.5 & 38.7 \\
Biomedical & 0.161 & 31.8 & 31.9 & 27.5 & 25.8 \\
Chemistry & 0.566 & 45.7 & 44.7 & 40.0 & 36.2 \\
CS & 1.20 & 57.9 & 57.1 & 53.8 & 54.3 \\
Engineering & 0.360 & 54.0 & 53.2 & 48.4 & 45.5 \\
Env./Earth & 1.32 & 58.6 & 58.9 & 55.3 & 51.2 \\
Materials & 0.665 & 57.0 & 56.1 & 50.3 & 47.0 \\
Physics & 2.63 & 68.1 & 66.8 & 61.4 & 59.3 \\
\textbf{Mean} &  & \textbf{52.4} & \textbf{51.9} & \textbf{47.3} & \textbf{44.7} \\
\midrule
\multicolumn{6}{l}{\textit{L2 (agenda, hierarchical $\gamma=10^{-2}$)}} \\
\midrule
Biology & 0.0115 & 18.0 & 18.2 & 15.1 & 12.3 \\
Biomedical & 0.0088 & 11.5 & 11.6 & 9.4 & 8.1 \\
Chemistry & 0.0111 & 14.6 & 14.4 & 11.5 & 9.1 \\
CS & 0.0279 & 27.3 & 27.8 & 25.0 & 23.4 \\
Engineering & 0.0096 & 19.2 & 19.4 & 16.3 & 13.8 \\
Env./Earth & 0.0236 & 24.4 & 24.4 & 21.1 & 17.5 \\
Materials & 0.0179 & 21.3 & 20.8 & 17.2 & 14.5 \\
Physics & 0.0462 & 32.2 & 32.2 & 26.9 & 22.6 \\
\textbf{Mean} &  & \textbf{21.0} & \textbf{21.1} & \textbf{17.8} & \textbf{15.2} \\
\bottomrule
\end{tabular}
\end{table}

\begin{figure}[h]
  \centering
  \includegraphics[width=0.7\linewidth]{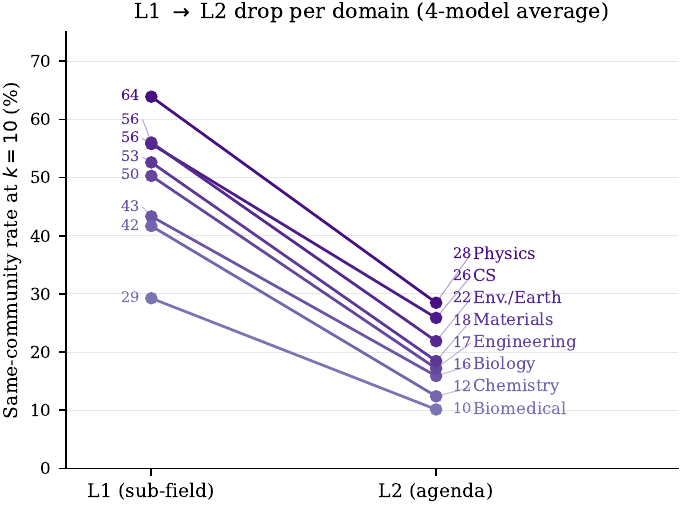}
  \caption{Per-domain L1$\to$L2 drop in top-10 same-community rate
  (four-model mean). Every domain loses 19 (Biomedical) to 35
  (Physics) percentage points when the partition tightens from
  sub-field to agenda granularity.}
  \label{fig:l1-l2-drop}
\end{figure}

The L1$\to$L2 drop is consistent across domains, ranging from 19
(Biomedical, which already starts low at L1, $29\%$) to 35
(Physics, $64 \to 28$) percentage points; the relative ordering of
domains is preserved (Figure~\ref{fig:l1-l2-drop}).

\subsection{The Gap Persists Across Neighbor Ranks}
\label{sec:results-rank}
\begin{figure}[h]
  \centering
  \includegraphics[width=\linewidth]{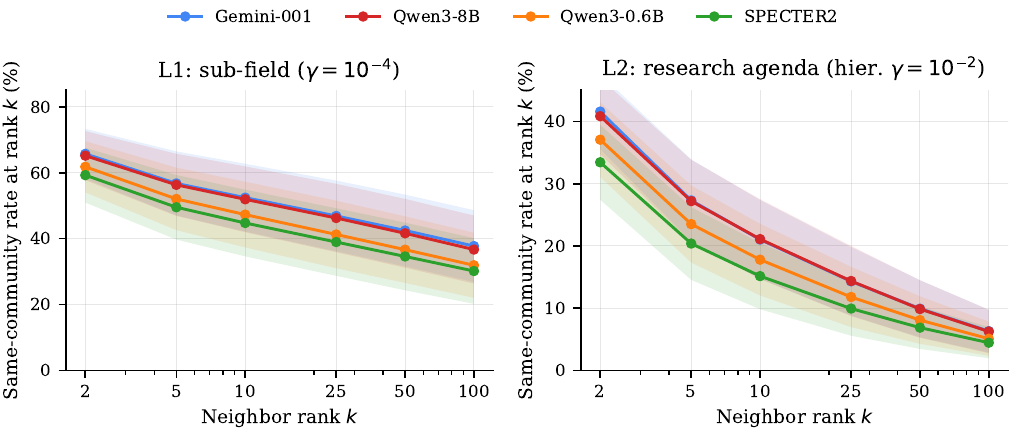}
  \caption{Same-community rate as a function of neighbor rank $k$,
  averaged across the eight domains. Shaded bands show $\pm 1$
  standard deviation across domains. Left: L1 (sub-field). Right:
  L2 (research agenda). The L2 curve decays faster than the L1
  curve at every model.}
  \label{fig:rank-sweep}
\end{figure}

Tightening top-$k$ does not rescue agenda match. At $k=2$ (the
single nearest neighbor) the L1 rate is 61--68\% and the L2 rate is
only 36--44\%; even the closest neighbor is wrong more than half the
time at the agenda level. At $k=100$ the L2 rate falls to 5--9\%.
The L1 curve decays gently with $k$ while the L2 curve decays much
faster, consistent with neighborhoods that broaden smoothly within
a sub-field but quickly fan out across research agendas. The
model ordering observed at $k=10$ (Section~\ref{sec:results-models})
holds at every $k$ we tested (Figure~\ref{fig:rank-sweep}).

\subsection{Model Ordering: SPECTER2 Is Weakest at Both Levels}
\label{sec:results-models}
The four models fall into a stable order across all eight domains
and both levels of community structure (Table~\ref{tab:hierarchical}):
$\text{Gemini} \,\approx\, \text{Qwen3-8B} \,>\,
\text{Qwen3-0.6B} \,>\, \text{SPECTER2}$. The Gemini--Qwen3-8B gap
is within 0.7\,pp at both levels, so the two leaders are effectively
tied. The ordering holds in every (model, domain) cell with one
exception: at L1 in Computer Science, SPECTER2 narrowly beats
Qwen3-0.6B ($54.3\%$ vs.\ $53.8\%$). Elsewhere the pattern is the
same.

Scaling from $0.6$B to $8$B buys 3--5 percentage points at L2, and
the remaining gap to Gemini is smaller still: parameter count
helps but does not close the sub-field--agenda gap. SPECTER2 uses
direct-citation pairs as positive contrastive examples, so its
cosine neighborhoods might be expected to track citation structure
downstream. In practice it underperforms the general-purpose
embeddings at both L1 ($44.7\%$ vs.\ $52.4\%$ for Gemini) and L2
($15.2\%$ vs.\ $21.0\%$). Training-time proximity on a subset of
citations does not, on its own, produce cosine neighborhoods that
match a carefully constructed citation partition.

\subsection{Domain Heterogeneity}
\label{sec:results-domain}
Concordance varies with domain (Table~\ref{tab:hierarchical}). The
L2 spread is nearly $3\times$: best is Physics--Gemini at
$32.2\%$, worst is Biomedical at $8.1$--$11.6\%$. The L2 ordering
of domains nearly matches the L1 ordering (Physics $>$ CS $>$
Env./Earth $>$ Materials $>$ Engineering $>$ Biology $>$ Chemistry
$>$ Biomedical). Physics and Env./Earth carry distinctive
instrument vocabulary that anchors a paper to a research
community; Biomedical and Chemistry reuse a small set of high-level
tokens (\emph{inhibitor}, \emph{receptor}, \emph{pathway}) across
mechanistically different agendas, so two papers on different
drugs can look almost identical in embedding space; CS is
in-between because task acronyms (FMEA, PSS, ELECTRE) plus an
application verb already approximate a named research thread.

\subsection{Manual Validation: What the L2 Gap Looks Like}
\label{sec:results-validation}
We manually inspected 16 representative queries (two per domain)
with their top-10 Qwen3-8B embedding neighbors to check that L2
communities correspond to what researchers would call ``the same
research agenda.'' Two illustrative cases:
a Physics query on UV \textsc{C\,iv} spectroscopy of Herbig Ae/Be
pre-main-sequence stars matches its L2 in only $1/10$ neighbors
($10/10$ at L1) -- all ten neighbors are Herbig Ae/Be papers, but
the other nine span Chandra X-ray, near-infrared Brackett-series,
GRAVITY interferometry, and SED modeling, distinct research
programs with non-overlapping foundational literature; a CS query
on smart-PSS FMEA matches its L2 in $8/10$ neighbors, the regime
where the query vocabulary is distinctive enough (task label +
methodology acronyms) for the embedding to collapse onto a single
agenda. The full 16-query set follows the same qualitative pattern
across all eight domains: distinctive agenda-identifying tokens
collapse the top-$k$ onto one L2, while topic-only tokens spread it
across several. Full data is in Appendix~\ref{app:l2-validation}.

\subsection{A Diagnostic Probe: Citation-Graph Rerank Recovers the Missing Signal}
\label{sec:results-retrieval}
If embeddings miss the L2 research agenda, does the augmented
graph itself carry the missing signal in retrievable form? We
treat this as a diagnostic probe rather than a proposed method:
the simplest citation-graph score (raw internal citation count)
applied on top of two off-the-shelf candidate generators (BM25
and LLM-expanded Boolean retrieval). The probe runs on 80 curated
agenda queries (10 per domain), each a short description of a
research thread plus a few representative papers. For each query
and retriever we report \emph{top-1 L2 same-rate}: the fraction
of queries whose rank-1 retrieved paper shares an L2 community
with at least one representative paper of the agenda.
We report seven retrievers in two groups: \emph{without citation
rerank} -- BM25 lexical retrieval and the four cosine
embeddings -- and \emph{with citation rerank} -- the citation-graph
retriever (LLM-expanded Boolean queries followed by internal
citation re-ranking, Section~\ref{sec:method-retrieval}) and a
BM25+cite variant that swaps the LLM-Boolean candidate generator
for plain BM25 but keeps the same citation rerank step.

\begin{table}[!htbp]
\centering
\footnotesize
\caption{Top-1 L2 same-rate ($\%$) on 80 curated research-agenda queries, across all retrievers. \emph{Without citation rerank} (BM25 lexical retrieval, four cosine models): top-1 L2 hovers at $39$--$51\%$. \emph{With citation rerank} (Graph: LLM-expanded Boolean queries followed by internal citation re-ranking; BM25+cite: BM25 retrieval followed by the same re-ranking step) both reach $\sim 58$--$60\%$, regardless of the candidate generator. RRF-based hybrid baselines we evaluated reach only 48--55\% (RRF over BM25+Gemini: 48\%; RRF over BM25+Gemini+citation: 55\%), suggesting naive score fusion is not the right way to combine these signals. Bold marks the per-domain winner.}
\label{tab:retrieval}
\begin{tabular}{lrrrrrrr}
\toprule
Domain & BM25 & SPECTER2 & Qwen3-0.6B & Qwen3-8B & Gemini & Graph & BM25+cite \\
\midrule
Biology & 21.4 & 26.6 & 20.6 & 36.3 & 37.9 & 53.0 & \textbf{60.0} \\
Biomedical & 24.9 & 32.8 & 23.5 & 19.1 & 34.5 & 38.0 & \textbf{42.0} \\
Chemistry & 40.8 & 50.3 & 45.8 & 46.2 & 55.0 & \textbf{63.0} & 60.0 \\
CS & 42.7 & 30.0 & 43.8 & \textbf{51.7} & 51.2 & 42.3 & 42.0 \\
Engineering & 43.0 & 48.2 & 43.4 & 35.9 & 45.3 & \textbf{62.0} & \textbf{62.0} \\
Env. Sci. & 34.2 & 38.8 & 41.6 & 49.0 & 45.4 & 56.0 & \textbf{68.0} \\
Materials & 38.0 & 41.7 & 35.6 & 48.5 & 51.9 & \textbf{68.0} & 62.0 \\
Physics & 69.0 & 49.2 & 62.9 & 72.4 & \textbf{84.0} & 79.0 & 81.0 \\
\midrule
\textbf{Mean} & 39.3 & 39.7 & 39.6 & 44.9 & 50.6 & 57.7 & \textbf{59.6} \\
\bottomrule
\end{tabular}
\end{table}

\begin{figure}[h]
  \centering
  \includegraphics[width=\linewidth]{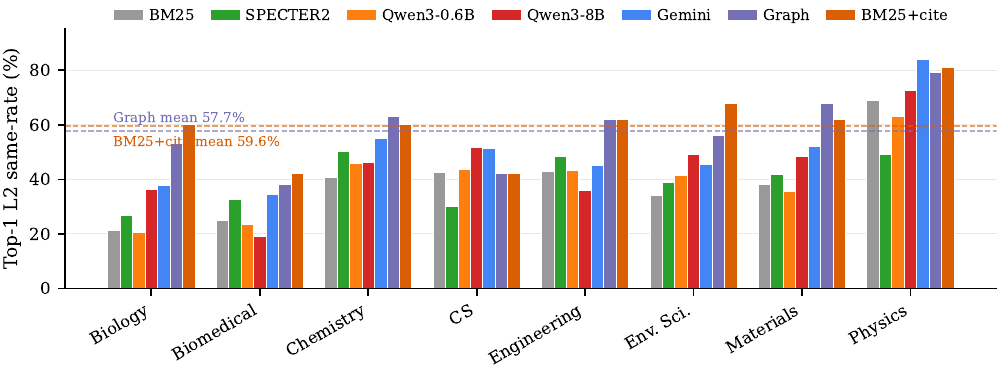}
  \caption{Top-1 L2 same-rate for the seven retrievers on 80
  curated research-agenda queries. Without citation rerank (BM25
  and the four cosine retrievers): the bars sit in the 39--51\%
  band. With citation rerank (Graph, BM25+cite): both clear all
  five candidate-only retrievers and end up within 2 points of
  each other. Dashed horizontal lines mark the 8-domain mean of
  Graph (57.7\%) and BM25+cite (59.6\%).}
  \label{fig:retrieval}
\end{figure}

Table~\ref{tab:retrieval} (visualised in Figure~\ref{fig:retrieval})
gives the result. The two groups separate
cleanly. Without citation rerank, BM25 (39.3\%), SPECTER2 (39.7\%),
Qwen3-0.6B (39.6\%), Qwen3-8B (44.9\%), and Gemini (50.6\%) all sit
in the 39--51\% band, with BM25 and the weaker dense models
indistinguishable and Gemini leading the cosine retrievers by
6 percentage points. With citation rerank, both Graph (57.7\%) and
BM25+cite (59.6\%) clear all five candidate-only retrievers and
end up within 2 points of each other. BM25+cite is the
per-domain winner in four of eight domains; Graph wins two; Gemini
and Qwen3-8B win one each (Physics and CS, the two domains where
surface text already carries the agenda).

Two observations follow. First, BM25 alone is competitive with the
two weaker dense retrievers (SPECTER2, Qwen3-0.6B) and well below
Gemini, which means that lexical retrieval is not the failure mode
either: a sparse retriever sits in the same 39--51\% band as the
cosine ones. The agenda-level miss is not specific to dense
embeddings. Second, what closes the gap is the citation-graph
re-ranking step, not the choice of candidate generator. Replacing
the LLM-expanded Boolean candidate set with raw BM25 candidates
preserves the gain (59.6\% vs.\ 57.7\%), so the contribution of our
pipeline is reproducible from a much simpler candidate generator.

The per-domain pattern matches what
Section~\ref{sec:results-domain} predicts. Citation-rerank wins
five of the six domains where the embedding cannot rely on
named-subfield surface vocabulary (Biology, Biomedical, Chemistry,
Engineering, Env.\ Sci., Materials), with margins of 3.5--16~pp.
The two embedding-favorable domains are Physics, where Gemini
(84\%) leads both citation-rerank methods, and CS, where Qwen3-8B
(51.7\%) does. These are the same domains in which agenda labels
(e.g., \emph{Graph Neural Networks $\&$ Graph Transformers},
\emph{Gravitational Wave Detection}) are themselves named
subfields heavily represented in modern web-scale embedding
training corpora; specialist Biomedical or Chemistry vocabulary is
comparatively rare in such corpora. We read the CS and Physics
advantage as a probable training-corpus bias on the embedding
side. Citation-rerank wins where embeddings cannot rely on it,
which is the regime covering most of scientific literature.

The diversity story matches: graph top-10 covers 4.4 distinct L2
communities on average versus 5.2--6.0 for the cosine retrievers,
so the cite-rerank top-10 is more agenda-cohesive. Three
Reciprocal Rank Fusion baselines combining BM25, Gemini, and
citation scores with fixed weights reach 47.9--55.3\% -- all
\emph{below} either citation-rerank method on its own, suggesting
naive score fusion dilutes the citation signal. The 7--9~pp gain
is a lower bound: the best retriever still misses 40\% of top-1
agendas, which we read as the same failure mode
Section~\ref{sec:results-validation} illustrates manually.

\section{Discussion}
\label{sec:discussion}

\subsection{Where Embeddings Stop Working}
Text embeddings are trained by co-occurrence; sub-field shares
that surface content abundantly, which is why L1 match at
45--52\% is non-trivial. Research agendas, in contrast, are
defined by a conjunction of specifics the text does not always
expose -- instruments, mechanisms, drugs, organisms, methods --
so two papers in the same L1 but different L2 look nearly
identical to a text encoder. SPECTER2 shows the limit of a
training-time fix: a direct-citation contrastive loss is a
topic-level signal, since agenda-level citations are only a
small slice of a paper's reference list. The augmented citation
graph instead accumulates agenda-level evidence across many
papers via bibliographic coupling and co-citation, which a
pair-level loss cannot match.

\subsection{Relation to Theoretical Limits}
\label{sec:disc-theory}
Our measurement gives a concrete empirical picture of the
expressivity bound proved by \citet{weller2026theoretical}: four
independently trained models converge to the same 15--21\% top-10
L2 same-rate, suggesting a representational wall rather than a
training-data artifact.

\subsection{Implications for Scientific RAG}
Our measurement explains a common practitioner-reported failure:
the top-$k$ pool is on the right topic but methodologically
distant. HyDE~\citep{gao2023hyde} and query-rewriting inherit the
same candidate pool with the same agenda-level miss; a standard
cross-encoder reranker can re-order the pool but cannot add papers
the dense retriever did not surface, since the failure happens at
pool-construction time. The diagnostic probe
(Section~\ref{sec:results-retrieval}) suggests bibliographic
coupling and co-citation over the augmented graph carry the
agenda-level slice cosine misses; the RRF baselines underperform
pure citation-rerank, hinting that fair fusion likely needs
per-query routing. Designing a learned reranker that uses this
signal, and verifying the effect in other reference graphs (legal,
patents, clinical guidelines), are left to future work.

\subsection{Limitations}
\textbf{(i)} Citation communities are a proxy for research-agenda
structure; our hard partition slightly over-counts disagreement
for cross-agenda papers. \textbf{(ii)} The Graph retriever's query expander and relevance
judge are both gemini-flash-lite~2.5, from the same model family
as Gemini Embedding; the BM25+cite variant uses no LLM at the
candidate stage, removes the confound, and reproduces the gain
(59.6\% top-1 L2), but a fully non-Gemini Graph configuration is
still future work. \textbf{(iii)} Our
10-of-25 query selection kept the agendas with the largest
Boolean-search result sets, biasing the benchmark toward
well-populated agendas. \textbf{(iv)} The measurement covers
scientific literature only; generalisation to other
reference-graph-bearing corpora needs empirical verification.
\textbf{(v)} The citation-graph rerank
(Section~\ref{sec:results-retrieval}) is a deliberately simple
diagnostic probe; designing a learned reranker that uses the same
graph is out of scope for this measurement paper.

\section{Conclusion}
\label{sec:conclusion}
We measured where text embeddings stop working at the granularity
that matters for scientific RAG: 8 of 10 top-10 retrieved papers
are off-agenda across four embeddings and eight domains; a trivial
citation-count rerank reaches 57.7\%--59.6\% top-1 L2, isolating
the missed signal. Graph, community labels, queries, and code are
released with this paper.

\bibliographystyle{plainnat}
\bibliography{references}

\appendix

\section{Full Rank-Sweep Tables}
\label{app:full-tables}
The full per-(model, domain, $k$) same-community rates and
enrichment ratios for $k \in \{2, 5, 10, 25, 50, 100\}$ at both L1
and L2 are released as JSON files alongside the paper.

\section{Embedding Configuration Details}
\label{app:embedding-config}
Per-model details for the title--abstract concatenation and pooling
reported in Section~\ref{sec:method-embeddings}:
\begin{itemize}
  \item \textbf{Gemini Embedding 001.} Title and abstract joined by a
        newline character; max input length 2048 tokens; last-token
        pooling via the public embedding API (output dimension 768).
  \item \textbf{Qwen3-Embedding-8B / 0.6B.} Title and abstract joined
        by a newline; max input length 512 tokens; last-token pooling
        on the final hidden state; $\ell_2$ normalization.
  \item \textbf{SPECTER2.} Title and abstract joined by the tokenizer's
        \texttt{[SEP]} token; max input length 512 tokens;
        \texttt{[CLS]} token pooling; $\ell_2$ normalization. We use
        the \texttt{allenai/specter2\_base} checkpoint without an
        adapter, matching the default usage reported
        by~\citet{singh2023scirepeval}.
\end{itemize}
All embeddings are $\ell_2$-normalized before indexing so that FAISS
inner-product search returns cosine similarity directly.

\section{Resolution Sensitivity Details}
\label{app:resolution}
We swept the Leiden CPM resolution parameter $\gamma$ across eight
values from $10^{-6}$ to $10^{-2}$ and chose
$\gamma_\text{L1}=10^{-4}$ for the L1 pass. At
$\gamma_\text{L1} \leq 10^{-6}$ a single catch-all community
absorbs roughly a quarter of the graph; at $\gamma_\text{L1} \geq
5\times 10^{-3}$ communities are too small to map onto recognisable
sub-fields. The intermediate values $5\times 10^{-5}$,
$10^{-4}$, and $5\times 10^{-4}$ all produce 70K--90K communities
with similar top-10 same-rate at the embedding side. We chose
$10^{-4}$ as the value at which manual inspection of the largest
clusters gave the cleanest sub-field labels. Per-$\gamma$ cluster
size statistics are released in
\texttt{leiden\_cpm\_sweep\_v2\_stats.json}.

\section{L2 Validation: 16-Query Manual Read}
\label{app:l2-validation}
We selected two queries per domain (16 total) and inspected the
top-10 Qwen3-8B embedding neighbors of each, comparing their L1
sub-field and L2 research-agenda labels with manual reading of the
title and abstract. Two of the 16 are summarised in
Section~\ref{sec:results-validation}; the full per-query labels,
neighbor titles, and L1/L2 agreement counts are released in
\texttt{subfield\_rerun\_hier.json} and
\texttt{knn\_hier\_discordance.json}.

\section{Why Domain Performance Varies (Lexical Analysis)}
\label{app:lexical}
Embeddings do best on CS and Physics, where graph offers little
extra gain. The naive explanation (more distinctive agenda
vocabulary) is inverted in the data: for each L1 sub-field with
multiple L2 communities, we computed the TF-IDF-weighted fraction
of unigrams that appear in only one of its L2 agendas. Aggregated
by domain, this fraction is highest in Biomedical ($22.2\%$),
Chemistry ($14.8\%$), and Biology ($10.8\%$) -- the three domains
with the lowest L2 agenda match
(Section~\ref{sec:results-domain}) -- and lowest in CS ($6.5\%$)
and Physics ($5.7\%$), which lead on agenda match.
We read this inversion as evidence that embeddings encode
\emph{concept composition} (\emph{transformer for vision} vs.\
\emph{transformer for language}) more readily than
\emph{entity-level} distinctions (\emph{doxorubicin} vs.\
\emph{paclitaxel}); citation graph signal naturally separates
entity-level lineages. A second factor is corpus bias: the CS and
Physics agenda labels are over-represented in modern embedding
training corpora, while specialist Biomedical or Chemistry
vocabulary is comparatively rare. The CS--Physics advantage is
therefore part-mechanistic and part-corpus-bias; graph wins in the
regime covering most of scientific literature. We caveat this
interpretation as post-hoc on eight domains.

\section{LLM Prompts}
\label{app:prompts}
The citation-graph retriever (Section~\ref{sec:method-retrieval})
uses two independent LLM agents, both running gemini-flash-lite 2.5
with no shared context. The exact system prompts are reproduced
below verbatim.

\subsection*{Search-Strategist Agent (Boolean Query Generation)}
\begin{quote}\small
You are an expert Search Strategist. Your mission is to analyze the
user's research intent and generate a diverse portfolio of 3--5
high-quality Boolean search queries for parallel execution. Your
goal is to design a set of queries that, when run simultaneously,
will provide a comprehensive and multi-faceted overview of the
topic.

\textbf{Strategic Portfolio Planning.} Based on a conceptual
analysis of the research goal, create a portfolio of 3--5 distinct
queries. Each query must be crafted to explore the topic from a
different strategic angle: at least one broad query to map the
landscape, at least one narrow and precise query to target the
most critical publications. Avoid creating simple variations of
the same query.

\textbf{Guiding Principles.} Use academic language (precise nouns,
established technical phrases; avoid buzzwords). Combine terms
hierarchically: \emph{Level~1} space-separated terms (default,
preferred); \emph{Level~2} \texttt{OR} within parentheses for
synonyms; \emph{Level~3} explicit \texttt{AND} for distinct concept
groups; \emph{Level~4} exact phrase match only for ambiguity
resolution or chemical/formal-entity distinction. Nested
parentheses are strictly prohibited; queries must be a flat series
of \texttt{AND}-connected groups.

\textbf{Output.} A single JSON object with key
\texttt{queries} holding the list of Boolean strings.
\end{quote}

\subsection*{Research-Analyst Agent (Relevance Scoring)}
\begin{quote}\small
You are a meticulous AI Research Analyst. Your function is to
process parallel academic search results and assign a
\texttt{relevance\_score} to each result based on the user's
research intent.

\textbf{Core principle.} For each search result, you score the
\emph{actual} \texttt{sample\_papers} titles, not the query text. A
query whose returned papers are off-topic is irrelevant regardless
of how the query reads.

\textbf{Score scale} (0.0--1.0):
\textbf{1.0}: nearly all sample paper titles are highly relevant;
\textbf{0.8--0.99}: most relevant with minor deviations;
\textbf{0.6--0.79}: a good portion relevant, noticeable noise;
\textbf{0.4--0.59}: about half relevant, search is struggling;
\textbf{0.0--0.39}: very few or no titles are relevant
(includes papers from completely different fields).

\textbf{Output.} A single JSON object with key
\texttt{evaluations} holding one object per input query, each with
\texttt{query}, \texttt{relevance\_score}, and a one-sentence
\texttt{analysis\_summary}.
\end{quote}

\section{Compute Resources}
\label{app:compute}
The pipeline runs on a single workstation (one consumer-grade
GPU, 22-core CPU, 80\,GB RAM for DuckDB out-of-core operations),
with one external step (Qwen3-8B inference on the 1.6M paper
pool) executed on a separate node with an NVIDIA A100 (80\,GB).
Approximate wall-clock cost per stage:
\begin{itemize}
  \item \textbf{Augmented graph construction.} DuckDB self-joins
        on 14.5\,GB of reference shards (16 threads, 80\,GB
        memory cap, disk spill enabled): bibliographic-coupling
        join $\sim 2$\,min, co-citation join $\sim 1$\,min,
        three-layer integration into 153M edges $\sim 3$\,min.
        Total compute under 10 minutes (excluding the 14.5\,GB
        shard download).
  \item \textbf{Leiden CPM community detection.} A single L1 run
        ($\gamma = 10^{-4}$) takes 65--80 minutes on one CPU
        worker via \texttt{leidenalg}. The L1 resolution sweep
        over 8 $\gamma$ values (Appendix~\ref{app:resolution})
        runs in $\sim 3$~hours with 4 parallel workers
        (\texttt{multiprocessing.Pool} with spawn context). The
        L2 hierarchical pass ($\gamma = 10^{-2}$) splits the
        graph into 1{,}896 per-L1 induced subgraphs and
        finishes in $\sim 4$ minutes with 6 parallel workers
        (subgraphs are small). \texttt{leidenalg} is CPU-only;
        no GPU acceleration is used at this stage.
  \item \textbf{Embedding inference for the 1.6M pool.} SPECTER2
        ($\sim 20$\,min) and Qwen3-0.6B ($\sim 30$\,min) run on
        the workstation GPU. Qwen3-8B requires more VRAM than
        the workstation provides and runs on a separate A100
        (80\,GB) node. Gemini Embedding 001 is queried via the
        Vertex AI API in batched JSONL chunks.
  \item \textbf{Top-$k$ neighbor analysis.} Four models
        $\times$ two community levels $\times$ eight domains
        $\times$ six $k$ values: GPU top-$k$ via PyTorch
        (5--10\,min/model) and CPU per-$k$ statistics
        ($\sim 30$\,min/model). Total $\sim 7$~hours wall-clock.
  \item \textbf{Retrieval evaluation (80 queries).} BM25
        indexing on 4M paper title\,+\,abstract via the
        \texttt{bm25s} library: $\sim 5$\,min total. The
        80-query cosine and citation-rerank passes run in
        under 1 minute each from cached embeddings and induced
        citation subgraphs. The Graph retriever uses
        gemini-flash-lite~2.5 for the Search-Strategist
        (3--5 calls/query) and the Research-Analyst (1 call per
        surviving candidate); $\sim 80 \times 6 = \sim 480$ API
        calls, $\sim 1$~hour wall-clock at default rate limits.
\end{itemize}
The reported pipeline fits in a single working day of wall-clock
time on a consumer-grade workstation, with one external A100 node
used for the Qwen3-8B paper-pool inference. Failed pilot
experiments -- a target-only graph variant whose CPM either
collapsed into a giant catch-all community or atomized at all
$\gamma$ values, ruled out before settling on the augmented graph;
alternative resolution sweeps; and abandoned hybrid scoring rules
-- added roughly 50\% on top of the reported pipeline time.

\section{Licenses for Existing Assets}
\label{app:licenses}
Data sources: OpenAlex bibliographic metadata
(\href{https://openalex.org/}{openalex.org}, CC0~1.0); PubMed
records (\href{https://pubmed.ncbi.nlm.nih.gov/}{public domain});
Semantic Scholar API (Allen Institute, S2 API terms of service).
For papers whose abstracts are missing across all three indexing
sources, a small fraction is retrieved from publisher websites
under fair-use research access and is \emph{not} redistributed
with our released artifacts. Embedding models: Gemini Embedding
001 (Google, accessed via API under
\href{https://ai.google.dev/terms}{Google AI API terms});
Qwen3-Embedding-8B and Qwen3-Embedding-0.6B (Alibaba,
\href{https://www.apache.org/licenses/LICENSE-2.0}{Apache 2.0});
SPECTER2 \texttt{allenai/specter2\_base} (AllenAI, Apache 2.0).
Libraries: \texttt{rank\_bm25} (MIT), \texttt{python-igraph}
(GPL-2.0+), \texttt{leidenalg} (GPL-3.0), \texttt{faiss}
(MIT), \texttt{sentence-transformers} (Apache 2.0), PyTorch (BSD-3).
The released artifacts (community labels, 80 curated queries,
analysis code) are licensed CC-BY 4.0 (data) and MIT (code), and
are mirrored as follows.

\paragraph{Code and analysis scripts.}
\url{https://github.com/junseon-yoo/topic-not-agenda}

\paragraph{Augmented citation graph and L1/L2 community labels.}
Zenodo, DOI \href{https://doi.org/10.5281/zenodo.20046263}{10.5281/zenodo.20046263}
(CC-BY~4.0).

\section{Curated Research-Agenda Queries}
\label{app:queries}
Table~\ref{tab:queries} lists the 80 curated agenda queries used
in Section~\ref{sec:results-retrieval}. Representative-paper IDs
for each query are released alongside the data files.

{\footnotesize
\begin{longtable}{p{0.92\linewidth}}
\caption{The 80 curated research-agenda queries used in the retrieval comparison (Section~\ref{sec:results-retrieval}). Ten queries per domain. Each query is a one-line description of a research thread.}\label{tab:queries}\\
\toprule
\textbf{Biology} \\
\midrule
\endfirsthead
\multicolumn{1}{l}{\textit{Table~\ref{tab:queries} continued from previous page.}} \\
\toprule
\endhead
\midrule
\multicolumn{1}{r}{\textit{Continued on next page.}} \\
\endfoot
\bottomrule
\endlastfoot
AI-based Medical Image Segmentation (COVID-19 \& pathology) \\
GAN-based Medical Image Augmentation \\
Bio-inspired Optimization Algorithms for Healthcare \\
Deep Learning for Neurological Disease Diagnosis (ASD \& MS) \\
Explainable \& Transparent AI for Medical Imaging \\
Remote Sensing for Insect \& Wildlife Monitoring \\
Insect Brain Architecture \& Neural Connectomics \\
Sex Determination Evolution \& Gender Biology \\
Mitochondrial Dynamics / Fusion-Fission \& ROS Production \\
Phylogenetic Network Theory \& Evolutionary Systematics \\
\midrule
\textbf{Biomedical} \\
\midrule
Medical Image Segmentation Algorithms (general frameworks) \\
Transformer Architectures in Medical Imaging \\
GAN-based Medical Image Augmentation \\
Domain Adaptation for Medical Image Analysis \\
Hydrogel-based Tissue Regeneration (collagen / fibrin / hyaluronic acid) \\
Decellularized Extracellular Matrix (ECM) Scaffolds \\
3D Bioprinting \& Bioink Engineering \\
Bone Tissue Engineering \& Hierarchical Scaffolds \\
Neural Tissue Injury Treatment (hydrogel-based approaches) \\
Nanomaterial-based Cancer Therapy (photothermal / ferroptosis / glutathione depletion) \\
\midrule
\textbf{Chemistry} \\
\midrule
Wearable \& Flexible Sensors (Sweat / Pressure / Pulse Wave) \\
MOF-based Porous / 2D Materials \& Enzyme Biocomposites \\
Electrocatalysis for Organic Synthesis \& Nitrogen Cycle \\
Covalent Organic Frameworks (COFs) for Electrochemical Applications \\
Photothermal Nanomaterials \& PDT-PTT Combination Therapy \\
Dynamic Covalent \& Unconventional Soft Polymer Networks \\
Direct Ink Writing \& 3D Printing of Diverse Materials \\
MXene: Synthesis Progress \& Functional Applications \\
Nanoparticle Endocytosis \& Targeting Strategies in Nanomedicine \\
Triboelectric Nanogenerators for Energy Harvesting \\
\midrule
\textbf{Computer Science} \\
\midrule
LLM Preference Optimization \& Reward Modeling \\
Transformer Architecture Innovations (Efficient / Hierarchical / MLP-based) \\
Few-Shot Visual Language Models \\
Semantic Segmentation with Transformers \\
Self-Supervised / Contrastive Representation Learning \\
Graph Neural Networks \& Graph Transformers \\
Multi-Agent \& Cooperative Reinforcement Learning \\
Video Understanding with Transformers \\
Time Series Forecasting with Transformers \\
Neural Implicit Surfaces \& 3D Reconstruction \\
\midrule
\textbf{Engineering} \\
\midrule
Bio-inspired \& Nature-inspired Metaheuristic Optimization Algorithms \\
Swarm Intelligence-based Feature Selection Methods \\
Fault Diagnosis with Small / Imbalanced Data (GAN \& Meta-learning) \\
Cross-domain \& Transfer Learning for Fault Diagnosis \\
Predictive Maintenance \& AI for Prognostics and Health Management (PHM) \\
Structural Health Monitoring with Machine Learning \\
Additive Manufacturing: Polymeric Composites \& Wire Arc (WAAM) \\
Natural Fiber Reinforced Composites for Sustainable Applications \\
Neural Network-based Constitutive Modeling of Composite Materials \\
Digital Twin for Product Design \& Engineering Systems \\
\midrule
\textbf{Environmental/Earth Sciences} \\
\midrule
Persulfate-based Nonradical \& Selective Oxidation Mechanisms \\
PFAS Occurrence / Transformation \& Soil-Water Remediation \\
Heavy Metal Removal from Water \& Wastewater \\
Microplastics Behavior \& Toxicity in Aquatic-Soil Systems \\
MOF-based Engineered Materials for Water Remediation \\
Circular Economy: Definitions / Critiques \& Measurement Indicators \\
Circular Economy Strategies for Climate Change \& Carbon Neutrality \\
Digital Technologies for Circular Economy (Blockchain \& Industry 4.0) \\
Biomass Conversion / Waste Recycling \& Municipal Solid Waste Management \\
Machine Learning \& Renewable Energy for Environmental Sustainability \\
\midrule
\textbf{Materials Science} \\
\midrule
Flexible MXene Composites for Wearable Devices \\
Direct Ink Writing / Metal Additive Manufacturing \& DED \\
Halide Perovskite Nanocrystals \& Optoelectronics \\
Hydrogel-based Flexible Electronics \\
S-Scheme \& Advanced Photocatalysts \\
Solid Electrolyte Interphase (SEI) \& Li Metal Anode Engineering \\
Composite Polymer Electrolytes for Solid-State Li Batteries \\
Electrolyte Solvation Structure \& Interfacial Modeling in Batteries \\
Water Splitting \& Hydrogen Evolution Reaction (HER) Catalysts \\
Electromagnetic Wave Absorption Materials \\
\midrule
\textbf{Physics} \\
\midrule
Variational Quantum Eigensolver (VQE) \& Optimization Methods \\
Tensor Network Methods (Matrix Product States / PEPS) \\
Quantum Heat Transport \& Finite-Temperature Dynamics \\
Non-Hermitian Physics \& Exceptional Topology \\
Primordial Black Holes \& Cosmological Constraints \\
Hubble Tension \& Cosmological Solutions \\
Teleparallel Gravity \& Modified Cosmology \\
Axion Dark Matter Detection \\
Rapid Neutron-Capture Process \& Heavy Element Origins (r-process) \\
Gravitational Wave Detection (space-based antennas) \\
\end{longtable}
}


\end{document}